\documentclass[12pt]{article}
\usepackage{amsmath,amssymb}
\begin{document}

\begin{titlepage}

\begin{flushright}
IZTECH-PH-2014/1
\end{flushright}

\bigskip

\begin{center}

{{\Large Riemann-Eddington theory:}\\

{\Large Incorporating matter, degravitating the cosmological constant}}

\bigskip

\bigskip

{\bf  Durmu{\c s} Ali Demir}\\
\smallskip

{ \small \it
Department of Physics, {\.I}zmir Institute of Technology, TR35430, {\.I}zmir, Turkey\\
Abdus Salam International Centre for Theoretical Physics, ICTP, I-34151, Trieste, Italy}

\bigskip

{\tt demir@physics.iztech.edu.tr}

\bigskip

\vspace*{.5cm}

{\bf Abstract}\\
\end{center}
\noindent
Here we show that, Eddington's pure affine gravity, when extended with Riemann curvature, 
leads to gravitational field equations that incorporate matter. This Riemanned Eddington gravity 
outfits a setup in which matter gravitates normally with Newton's constant but vacuum 
gravitates differently with an independent gravitational constant. This novel setup 
enables degravitation of the vacuum to observed level not by any fine-tuning but by a 
large hierarchy between its gravitational constant and its energy density. Remarkably, 
degravitation of the cosmological constant is local, causal and natural yet only empirical 
because the requisite degravitation condition is not predicted by the theory.
\\ 
\bigskip

\noindent PACS: 04.50.-h, 04.20.Cv, 95.30.Sf  

\bigskip

\end{titlepage}

\section{Introduction}
Soon after the foundation of General Relativity (GR) \cite{einstein1} as the
theory of gravitation, attempts to generalize it were started (see the reviews
\cite{review} and \cite{review2}). One such attempt was the introduction of the cosmological constant \cite{einsteincc}, which
initiated the modern cosmology. Another attempt of great physical and mathematical
interest was Eddington's affine gravity \cite{eddington1,eddington2}. His idea was to 
use the affine connection itself as the fundamental quantity to describe gravitation.
This proposal was furthered first by Einstein \cite{einstein2,einstein3} and later
by Schroedinger \cite{schroedinger1} (see also the solutions in \cite{schroedinger2}).

In general, the problem with Eddington's approach is that the Einstein equations it yields 
involve the cosmological constant and only the cosmological constant. The equations exhibit no sensitivity
to matter and radiation in the Universe. This means that matter does not gravitate,
and all bodies stay weightless in this model. The model must be improved to avoid this unphysical 
situation. For this purpose, Banados and Ferreira \cite{banados1} extended the  Eddington theory by forcibly 
adding matter action to it (in spacetimes endowed with metric). This proposal has 
resulted in revival of Eddington's theory with various cosmological and astrophysical applications. 
Nevertheless, this ``Eddington-inspired Born-Infeld gravity'' has been found to suffer 
from instabilities \cite{banados2} (which, according to \cite{banados2p}, might be resolved by 
back-reaction of particles). In any case, coupling of gravity to matter is not a 
settled issue in Eddington formalism, and it needs a clear resolution before 
studying its cosmological, astrophysical and other implications. 

In the present work we tackle this problem. We show that Eddington's approach
leads to correct gravitational field equations if it is extended to involve the grand
curvature of spacetime, encoded in the Riemann tensor. Our setup has its roots in the
Riemann-only gravitational theory constructed by the author and his collaborators
in \cite{non-lin2}. In that work, the Einstein field equations were extracted from
rank-4 dynamical equations by expressing torsion contribution as the 
Kulkarni-Numizu product of the metric and energy-momentum tensors. In general, 
affine gravity provides an appropriate framework for a dynamical approach 
to metrical theories of gravity \cite{review2,non-lin,non-lin2}. 
The setup of the present work, the ``Riemann-improved Eddington theory'' as we will 
call it from now on, completes Eddington theory by incorporating the missing matter 
terms properly (by inducing a general energy-momentum tensor generically 
valid for all kinds of matter fields). Interestingly, this improved theory not 
only induces matter dynamically but also enables degravitation of the 
cosmological constant in a local, causal and not fine-tuned way. Stating differently, in 
the consequent gravitational theory, matter and vacuum gravitate with different 
gravitational constants, and bounds on cosmological constant are 
satisfied not by fine-tuning but by large hierarchies. These sensible aspects do not, however, mean that 
the cosmological constant problem is solved. The reason is that degravitation 
of the vacuum energy necessitates fundamental parameters to be related to the 
vacuum energy in a specific way, and the model is unable to offer any dynamical
mechanism that predicts the aforementioned relation. The said relation is thus 
empirical rather than fundamental.
 
In Sec. II below we briefly review the Eddington theory. In Sec. III, we study in 
detail construction, dynamics, and gravitational properties or matter and vacuum 
within the Riemann-improved Eddington theory. In Sec. IV we conclude.

\section{Eddington Gravity}
Despite its unphysical results, Eddington's idea is to the point. The
reason is that it is based exclusively on the affine connection $\Gamma^{\mu}_{\alpha\beta}\,$,
which is a purely geometrical acceleration field that fully governs the spacetime curvature.
As a matter of the fact, the Riemann tensor
\begin{eqnarray}
\label{riemann}
{\mathfrak{R}}^{\mu}_{\alpha\nu\beta}\left(\Gamma\right) = \partial_{\nu} \Gamma^{\mu}_{\beta\alpha} - \partial_{\beta} \Gamma^{\mu}_{\nu\alpha}
+\Gamma^{\mu}_{\nu\lambda} \Gamma^{\lambda}_{\beta\alpha} - \Gamma^{\mu}_{\beta\lambda} \Gamma^{\lambda}_{\nu\alpha}
\end{eqnarray}
involves only connection and constitutes a unique measure of the spacetime curvature in that spacetime is flat
if and only if all components of ${\mathfrak{R}}^{\mu}_{\alpha\nu\beta}\left(\Gamma\right)$ vanish.
It contracts to produce the Ricci tensor
\begin{eqnarray}
\label{ricci}
{\mathcal{R}}_{\alpha\beta}\left(\Gamma\right) \equiv {\mathfrak{R}}^{\mu}_{\alpha\mu\beta}\left(\Gamma\right) =
\partial_{\mu} \Gamma^{\mu}_{\beta\alpha} - \partial_{\beta} \Gamma^{\mu}_{\mu\alpha}
+\Gamma^{\mu}_{\mu\lambda} \Gamma^{\lambda}_{\beta\alpha} - \Gamma^{\mu}_{\beta\lambda} \Gamma^{\lambda}_{\mu\alpha}
\end{eqnarray}
which covers only a subset of the grand curvature components in (\ref{riemann}). Its antisymmetric part ${\mathcal{R}}_{[\alpha\beta]}$
is given by ${\mathfrak{R}}^{\mu}_{\mu\alpha\beta}\left(\Gamma\right)$. The Weyl
curvature tensor ${\mathfrak{W}}^{\mu}_{\alpha\nu\beta}\left(\Gamma\right)$ is completely traceless and enjoys the same symmetries 
as the Riemann tensor. 

For a symmetric connection, $\Gamma^{\mu}_{\alpha\beta} = \Gamma^{\mu}_{\beta\alpha}$,
as also assumed in Eddington's original proposal \cite{eddington1}, torsion vanishes and curvature
remains as the only tensorial object in affine spacetime \cite{affine}. Having no metric tensor,
curvature tensors  cannot be contracted any further to obtain scalars (like curvature scalar in GR). The absence
of scalars, however, does not obstruct the construction of invariant actions because
differential volume element $d^4 x$ transforms not as a scalar but as a scalar density
under coordinate mappings, and therefore, all one needs is a scalar density ${\mathcal{L}}$
such that the product $d^{4}x\ {\mathcal{L}}$ is a scalar. Generically, the said
${\mathcal{L}}$ is provided by determinants of tensors. The most obvious
choice is the Ricci tensor, and it leads to the Eddington action \cite{eddington1,eddington2}
\begin{eqnarray}
\label{ac-eddington}
I_{E}\left[\Gamma\right] = \int d^{4}x \sqrt{{\texttt{Det}}\left[{\mathcal{R}}\left(\Gamma\right)\right]}
\end{eqnarray}
wherein the determinant of ${\mathcal{R}}_{\alpha\beta}$
\begin{eqnarray}
\label{det-ricci}
{\texttt{Det}}\left[{\mathcal{R}}\right] = \frac{1}{4 !}\epsilon^{\alpha_0 \alpha_1 \alpha_2 \alpha_3} \epsilon^{\beta_0 \beta_1 \beta_2 \beta_3}
{\mathcal{R}}_{\alpha_0\beta_0}{\mathcal{R}}_{\alpha_1\beta_1}{\mathcal{R}}_{\alpha_2\beta_2}{\mathcal{R}}_{\alpha_3\beta_3}
\end{eqnarray}
involves only the antisymmetric Levi-Civita symbol $\epsilon^{\alpha_0 \alpha_1 \alpha_2 \alpha_3}$. Stationarity of the
action (\ref{ac-eddington}) against the variation
\begin{eqnarray}
\label{variation}
\delta {\mathfrak{R}}^{\mu}_{\alpha\nu\beta} = \nabla_{\nu}\Big( \delta\Gamma^{\mu}_{\beta\alpha}\Big) -
\nabla_{\beta}\Big(\delta\Gamma^{\mu}_{\nu\alpha}\Big)
\end{eqnarray}
gives the equation of motion
\begin{eqnarray}
\label{eom-eddington2}
\nabla_{\mu} \left[ \sqrt{{\texttt{Det}}\left[{\mathcal{R}}\right]}\left({\mathcal{R}}^{-1}\right)^{\alpha\beta}\right]=0
\end{eqnarray}
which is solved by
\begin{eqnarray}
\label{solution0}
\sqrt{{\texttt{Det}}\left[{\mathcal{R}}\right]}\left({\mathcal{R}}^{-1}\right)^{\alpha\beta}  = \Lambda
\sqrt{g} g^{\alpha\beta}
\end{eqnarray}
provided that $g_{\alpha\beta}$ is an invertible, covariantly-constant tensor field with determinant $g={\texttt{Det}}\left[{g}_{\alpha\beta}\right]$
and inverse $g^{\alpha\beta} = \left(g^{-1}\right)^{\alpha\beta}$. The equality (\ref{solution0}) fines down to
\begin{eqnarray}
\label{solution}
{\mathcal{R}}_{\alpha\beta}\left(\Gamma\right) = \Lambda g_{\alpha\beta}
\end{eqnarray}
which is nothing but the Einstein field equations with cosmological constant $\Lambda$. This is so because $g_{\alpha\beta}$ has every reason
to qualify as the metric tensor on spacetime, and its covariant constancy, $\nabla_{\mu} g_{\alpha\beta} = 0$, fully determines the connection
$\Gamma^{\lambda}_{\alpha\beta}$ to be the Levi-Civita connection
\begin{eqnarray}
\label{LC}
{}^{g}\Gamma^{\lambda}_{\alpha\beta} = \frac{1}{2} g^{\lambda\rho} \left( \partial_{\alpha} g_{\beta\rho} +
\partial_{\beta} g_{\rho\alpha} - \partial_{\rho} g_{\alpha\beta}\right)
\end{eqnarray}
used in  GR \cite{einstein1}. In summary, Eddington's approach dynamically gives a GR-like setup in which
{\it (i)} Eddington's affine spacetime is dynamically endowed with a metric tensor and hence with the notions of distance and angle,  {\it (ii)} affine
connection converts into the Levi-Civita connection in GR, and {\it (iii)} Einstein field equations in GR are only partially reproduced
because the right-hand side of (\ref{solution}) lacks the energy-momentum tensor $T_{\alpha\beta}$ of matter. This is a disaster.
In recent years, phenomenological attempts have been made to improve it by adding to (\ref{ac-eddington}) the matter action directly ``$\dots$
without insisting neither on a purely affine action nor on a theory equivalent to Einstein gravity." \cite{banados1}. This model has important
astrophysical and cosmological implications \cite{eddington-astro-cosmo} yet it has been found \cite{banados2} to suffer from certain tensor
instabilities. In affine geometry, coupling of gravity to matter is a perplexing problem \cite{matter-grav}, and below we study it 
starting from first principles and staying parallel to Eddington's construction.

\section{Riemanning Eddington Gravity}
In this section we show that problems with Eddington's theory can be naturally resolved if it is extended to involve the grand curvature of spacetime. To this end, we extend the Eddington 
action (\ref{ac-eddington}) as 
\begin{eqnarray}
\label{ac-riemann} I_{R}\left[\Gamma\right] =
\int d^{4}x \left\{ a \sqrt{{\texttt{Det}}\left[{\mathcal{R}}\left(\Gamma\right)\right]} + b
\sqrt{{\texttt{Det}}\left[{\mathfrak{R}}\left(\Gamma\right)\right]}\right\}
\end{eqnarray}
where $a$, $b$ are dimensionless constants. They are true constants since in affine spacetime there exist no curvature invariants that 
can promote $a$ and $b$ to dynamical variables. In (\ref{ac-riemann}), determinant of the Riemann tensor
\begin{eqnarray}
\label{det-riemann}
{\texttt{Det}}\left[{\mathfrak{R}}\right] = \frac{1}{\left(4 !\right)^2}\epsilon_{\mu_0 \mu_1 \mu_2 \mu_3} \epsilon^{\alpha_0
\alpha_1 \alpha_2 \alpha_3} \epsilon^{\nu_0 \nu_1 \nu_2 \nu_3} \epsilon^{\beta_0 \beta_1 \beta_2 \beta_3}
{\mathfrak{R}}^{\mu_0}_{\alpha_0\nu_0\beta_0}{\mathfrak{R}}^{\mu_1}_{\alpha_1\nu_1 \beta_1} {\mathfrak{R}}^{\mu_2}_{\alpha_2\nu_2\beta_2}{\mathfrak{R}}^{\mu_3}_{\alpha_3\nu_3\beta_3}
\end{eqnarray}
generalizes the usual determinant in (\ref{det-ricci}) to rank-four tensors \cite{gelfand,deter1}. There are four $\epsilon$ symbols here yet 
the covariant--looking and contravariant--looking symbols contract as
$\displaystyle{\epsilon_{\mu \alpha \nu \beta} \epsilon^{\mu' \alpha'
\nu' \beta'} = \delta_{\left[\mu\right.}^{\left[\mu'\right.}
\cdots \delta_{\left.\beta\right]}^{\left.\beta'\right]}}$
to leave only two contravariant--looking $\epsilon$ symbols as in the determinant of the Ricci tensor in equation 
(\ref{det-ricci}).  Thus, Ricci and Riemann pieces in action (\ref{ac-riemann}) are tensor densities of identical 
weights, and hence, the Riemann piece also leads to an invariant action like (\ref{ac-eddington}).  The action
(\ref{ac-riemann}) stays stationary against the variation in (\ref{variation}) provided that 
\begin{eqnarray}
\label{eom-riemann}
\nabla_{\nu} \left[ b \sqrt{{\texttt{Det}}\left[{\mathfrak{R}}\right]}\left({\mathfrak{R}}^{-1}\right)^{\beta\nu\alpha}_{\quad \mu}
+ \frac{a}{2} \sqrt{{\texttt{Det}}\left[{\mathcal{R}}\right]}\left( \delta^{\nu}_{\mu} \left({\mathcal{R}}^{-1}\right)^{\beta\alpha}
-  \left({\mathcal{R}}^{-1}\right)^{\nu\alpha}\delta^{\beta}_{\mu}\right) \right] = 0
\end{eqnarray}
holds. In here, the inverse Riemann tensor \cite{deter1} 
\begin{eqnarray}
\label{inv-riemann}
\left({\mathfrak{R}}^{-1}\right)^{\beta\nu\alpha}_{\quad \mu} = \frac{1}{(3!)^2} \frac{1}{{\texttt{Det}}\left[{\mathfrak{R}}\right]} 
\epsilon_{\mu \mu_1 \mu_2 \mu_3} \epsilon^{\beta \beta_1 \beta_2 \beta_3} \epsilon^{\nu \nu_1 \nu_2 \nu_3}
\epsilon^{\alpha \alpha_1 \alpha_2 \alpha_3} {\mathfrak{R}}^{\mu_1}_{\alpha_1\nu_1 \beta_1} {\mathfrak{R}}^{\mu_2}_{\alpha_2\nu_2\beta_2}{\mathfrak{R}}^{\mu_3}_{\alpha_3\nu_3\beta_3}
\end{eqnarray}
is direct generalization of $\left({\mathcal{R}}^{-1}\right)^{\alpha\beta}$ to rank-4 tensor fields. It arises in (\ref{eom-riemann})
for the same reason that the inverse Ricci tensor $\left({\mathcal{R}}^{-1}\right)^{\alpha\beta}$ arises in (\ref{eom-eddington2}). The inverse
Riemann satisfies not only the usual relations like $\left({\mathfrak{R}}^{-1}\right)^{\beta\nu\alpha}_{\quad \mu}  {\mathfrak{R}}^{\mu'}_{\alpha\nu\beta} = \delta^{\mu'}_{\mu}$
which apply to rank-2 tensors as well but also the matrix multiplications like $\left({\mathfrak{R}}^{-1}\right)^{\beta\nu\alpha}_{\quad \mu}  
{\mathfrak{R}}^{\mu}_{\alpha\nu'\beta'} = \frac{1}{3} \left( \delta^{\nu}_{\nu'} \delta^{\beta}_{\beta'} - \delta^{\nu}_{\beta'} \delta^{\beta}_{\nu'}\right)$ 
which arise from its higher-rank nature.

Now, we are in a position to analyze the affine equation of motion (\ref{eom-riemann}). It can be integrated to find 
\begin{eqnarray}
\label{soln-riemann}
b \sqrt{{\texttt{Det}}\left[{\mathfrak{R}}\right]}\left({\mathfrak{R}}^{-1}\right)^{\beta\nu\alpha}_{\quad \mu}
+ \frac{a}{2} \sqrt{{\texttt{Det}}\left[{\mathcal{R}}\right]}\left( \delta^{\nu}_{\mu} \left({\mathcal{R}}^{-1}\right)^{\beta\alpha}
-  \left({\mathcal{R}}^{-1}\right)^{\nu\alpha}\delta^{\beta}_{\mu}\right) = {\mathfrak{T}}^{\beta\nu\alpha}_{\quad \mu}
\end{eqnarray}
where ${\mathfrak{T}}^{\beta\nu\alpha}_{\quad \mu}$ is a tensor density having the same symmetries 
as $\left({\mathfrak{R}}^{-1}\right)^{\beta\nu\alpha}_{\quad \mu}$. It is an integral of motion, and 
possesses the following properties:
\begin{flalign}
\label{cond1}
&\text{${\mathfrak{T}}^{\beta\nu\alpha}_{\quad \mu}$ must have vanishing divergence. This is because (\ref{soln-riemann}) gives back the affine}\nonumber\\
&\text{equation of motion  (\ref{eom-riemann}) provided that $\nabla_{\nu}{\mathfrak{T}}^{\beta\nu\alpha}_{\quad \mu} = 0$.}
\end{flalign}
\begin{flalign}
\label{cond2}
&\text{Being the integral of the equation of motion (\ref{eom-riemann}), ${\mathfrak{T}}^{\beta\nu\alpha}_{\quad \mu}$ cannot involve the}\nonumber\\
&\text{curvature tensors themselves, that is, it must be independent of ${\mathfrak{R}}^{\mu}_{\alpha\nu\beta}$, ${\mathcal{R}}_{\alpha\beta}$,}\\ 
&\text{${\mathfrak{W}}^{\mu}_{\alpha\nu\beta}$ and their contractions. By the same token, possible parameters in ${\mathfrak{T}}^{\beta\nu\alpha}_{\quad \mu}\;\;\;$}\nonumber\\ 
&\text{must be independent of the fundamental constants $a$ and $b$. It is through}\nonumber\\
&\text{the observational constraints that they can have a relation.}\nonumber
\end{flalign}
\begin{flalign}
\label{cond3}
&\text{The equation of motion (\ref{eom-riemann}) is a vanishing divergence. This means that ${\mathfrak{T}}^{\beta\nu\alpha}_{\quad \mu}$ is}\nonumber\\ 
&\text{not necessarily a function only of the metric tensor as in Eddington theory;}\\ 
&\text{it can involve novel tensor structures completely different to the metric tensor. }\nonumber
\end{flalign}
These are the primary conditions to be fulfilled by the solution of the equations of motion. Being a rank-4 tensor equation 
with no clue about the content of ${\mathfrak{T}}^{\beta\nu\alpha}_{\quad \mu}$, the affine equation of motion (\ref{soln-riemann}) 
can possess a variety of solutions characterized by the spacetime structures formed and spin and mass representations 
propagated. We, however, specialize here to one particular solution: The solution which leads to an equation for Ricci tensor in ways 
similar to the GR. In other words, we extract rank-2 portion of the rank-4 tensor equation (\ref{soln-riemann}) for the 
purpose of obtaining GR-like dynamics. In accordance with this,  in light of the three conditions above and
in the philosophy of (\ref{solution0}) and (\ref{solution}), we structure the motion integral ${\mathfrak{T}}^{\beta\nu\alpha}_{\quad \mu}$ 
by proposing the ansatz
\begin{eqnarray}
\label{soln2-riemann}
{\mathfrak{T}}^{\beta\nu\alpha}_{\quad \mu} &=& \sqrt{g} \bigg\{ \frac{\Lambda}{2} \left(\delta^{\nu}_{\mu} g^{\beta\alpha} - g^{\nu\alpha} \delta^{\beta}_{\mu}\right)
+ Q^{\beta\nu\alpha}_{\quad \mu}\nonumber\\
&+& c_1 \big( Q^{\beta\alpha} \delta^{\nu}_{\mu} - \delta^{\beta}_{\mu} Q^{\nu\alpha}\big)
+ c_2 \big( g^{\beta\alpha} Q^{\nu}_{\mu} - Q^{\beta}_{\mu} g^{\nu\alpha}\big)
+ c_3 Q \big( g^{\beta\alpha} \delta^{\nu}_{\mu} - \delta^{\beta}_{\mu} g^{\nu\alpha}\big)  \bigg\}
\end{eqnarray}
where $Q^{\beta\nu\alpha}_{\quad \mu}$ contains the non-metrical tensor structure mentioned in (\ref{cond3}).
It contracts to yield the lower-rank tensor fields $Q^{\beta\alpha} = Q^{\beta\nu\alpha}_{\quad \nu}$, $Q^{\nu}_{\mu} = 
g_{\beta\alpha}Q^{\beta\nu\alpha}_{\quad \mu}$, and $Q=g_{\beta\alpha} Q^{\beta\alpha}$. 

The  ansatz for ${\mathfrak{T}}^{\beta\nu\alpha}_{\quad \mu}$ in (\ref{soln2-riemann}) is composed of a gradient-free part proportional to $\Lambda$
and a divergence-free part involving $Q^{\beta\nu\alpha}_{\quad \mu}$. For this classification to make sense, it is essential that $Q^{\beta\nu\alpha}_{\quad \mu}$
does not contain any gradient-free part like the $\Lambda$ term in (\ref{soln2-riemann}). This constraint is a restatement of the property (\ref{cond3}). 

In regard to the property (\ref{cond1}), the divergence of ${\mathfrak{T}}^{\beta\nu\alpha}_{\quad \mu}$ must vanish. This 
does not imply a similar condition for $Q^{\beta\nu\alpha}_{\quad \mu}$. 
Indeed, it is possible that $\nabla_{\lambda} Q^{\beta\nu\alpha}_{\quad \mu} \neq 0$ yet its derivatives can organize to
precipitate $\nabla_{\nu}{\mathfrak{T}}^{\beta\nu\alpha}_{\quad \mu} = 0$. In fact, as emphasized
in (\ref{cond2}), ${\mathfrak{T}}^{\beta\nu\alpha}_{\quad \mu}$ must be independent of the curvature tensors, and 
this necessitates $Q^{\beta\nu\alpha}_{\quad \mu}$ to be also independent of the curvature tensors. This independence is ensured by making sure 
that $Q^{\beta\nu\alpha}_{\quad \mu}$ possesses certain properties which decisively distinguish it from curvature tensors.
To this end, Bianchi identities prove pivotal in that $Q^{\beta\nu\alpha}_{\quad \mu}$ is guaranteed to differ 
from the Riemann tensor if it enjoys, for instance, a differential identity of the form
\begin{eqnarray}
\label{diverg0}
\nabla_{\nu} Q^{\beta\nu\alpha}_{\quad \mu} =  c_4 \nabla_{\mu} Q^{\beta\alpha} + c_5 \nabla^{\alpha} Q^{\beta}_{\mu} + c_6 \nabla^{\beta} Q^{\alpha}_{\mu}
\end{eqnarray}
with not all of $c_4$, $c_5$, $c_6$ taking the values required by the Bianchi identities for curvature tensors. This Bianchi-inspired identity is an 
ansatz for $Q^{\beta\nu\alpha}_{\quad \mu}$. It gives 
\begin{eqnarray}
\label{diverg}
\nabla_{\nu} Q^{\nu}_{\mu} = \frac{c_4}{1-c_5-c_6} \nabla_{\mu} Q
\end{eqnarray}
upon contraction. Then, the identities (\ref{diverg0}) and (\ref{diverg}) facilitate the desired
relation $\nabla_{\nu} {\mathfrak{T}}^{\beta\nu\alpha}_{\mu} =  0$ for a generic $Q^{\beta\nu\alpha}_{\quad \mu}$
if $c_i$ satisfy the constraints
\begin{eqnarray}
\label{coeffs}
c_2 = -c_4 = c_5 = c_1,\, c_3 = \frac{c_1^2}{1-c_1},\, c_6 = 0
\end{eqnarray}
which give a realization of the property of ${\mathfrak{T}}^{\beta\nu\alpha}_{\quad \mu}$ given in (\ref{cond1}). 
The identity (\ref{diverg}) ensures that $Q^{\beta\nu\alpha}_{\quad \mu}$ vanishes identically 
unless $c_1 \neq 1$. With $ c_5 = - c_4 = c_1 \neq 1$, however, the identity (\ref{diverg0}) never conforms to 
the Bianchi identities for curvature tensors. As a result, the motion integral 
${\mathfrak{T}}^{\beta\nu\alpha}_{\quad \mu}$ is divergence-free as required by (\ref{cond1}), does not involve curvature tensors as 
required by (\ref{cond2}), and consists of non-metrical structures $Q^{\beta\nu\alpha}_{\quad \mu}$ 
as required by (\ref{cond3}). In conclusion, the ansatz (\ref{soln2-riemann}) for ${\mathfrak{T}}^{\beta\nu\alpha}_{\quad \mu}$
supplemented by the ansatz (\ref{diverg0}) for $Q^{\beta\nu\alpha}_{\quad \mu}$ yields a self-consistent picture in which 
all three conditions (\ref{cond1}), (\ref{cond2}) and (\ref{cond3}) are satisfied. This picture accommodates a divergence-free tensor field
\begin{eqnarray}
\label{Q-tensor}
K_{\alpha\beta} = Q_{\alpha\beta} + \frac{c_1}{1-c_1} Q g_{\alpha\beta}
\end{eqnarray}
as directly follows from (\ref{diverg}). This tensor field is strictly conserved, $\nabla^{\alpha} K_{\alpha\beta} = 0$. It
is guaranteed to involve no curvature contamination by (\ref{coeffs}). Thus, as will be shown below, it acts as a nontrivial source 
generalizing the homogeneous source $\Lambda g_{\alpha\beta}$ in the Eddington equation (\ref{solution}). 

Having constructed  ${\mathfrak{T}}^{\beta\nu\alpha}_{\quad \mu}$ and established the non-geometrical nature of 
$Q^{\beta\nu\alpha}_{\quad \mu}$, it is now time to solve for the curvature tensors themselves from (\ref{soln-riemann}). 
This is not so direct, however. The reason is that the left-hand side of (\ref{soln-riemann}) 
involves the inverse Riemann and inverse Ricci pieces together, and determining one of them necessitates a 
proper knowledge of the other. Thus it is with a physical ansatz and self-consistent 
determination of model parameters that one can extract the curvature tensors. Each of the two pieces, 
not a motion integral individually, violates all of (\ref{cond1}), (\ref{cond2}) and (\ref{cond3}). 
The inverse Riemann piece, for instance, does not have to have vanishing divergence. Also, 
it can explicitly involve the Weyl curvature (but not the Riemann and Ricci curvatures). 
Then, guided by the form of ${\mathfrak{T}}^{\beta\nu\alpha}_{\quad \mu}$ in (\ref{soln2-riemann}), 
it can be structured as
\begin{eqnarray}
\label{soln3-riemann}
\sqrt{{\texttt{Det}}\left[{\mathfrak{R}}\right]}\left({\mathfrak{R}}^{-1}\right)^{\beta\nu\alpha}_{\quad \mu} &=& \sqrt{g}\Big\{  {\mathfrak{W}}^{\beta\nu\alpha}_{\quad \mu} + 
 {\tilde{\Lambda}} \left(\delta^{\nu}_{\mu} g^{\beta\alpha} - g^{\nu\alpha} \delta^{\beta}_{\mu}\right)
+ Q^{\beta\nu\alpha}_{\quad \mu}
+ \tilde{c}_1 \big( Q^{\beta\alpha} \delta^{\nu}_{\mu} - \delta^{\beta}_{\mu} Q^{\nu\alpha}\big)\nonumber\\
&+& \tilde{c}_2 \big( g^{\beta\alpha} Q^{\nu}_{\mu} - Q^{\beta}_{\mu} g^{\nu\alpha}\big)
+   \tilde{c}_3 Q \big( g^{\beta\alpha} \delta^{\nu}_{\mu} - \delta^{\beta}_{\mu} g^{\nu\alpha}\big)  \Big\}
\end{eqnarray}
where the cofficients $\tilde{c}_i$ are different from $c_i$ in (\ref{soln2-riemann}) and do not have to take 
the values in (\ref{coeffs}). This is yet another ansatz. In reality, this ansatz does not have to depend on $Q^{\beta\nu\alpha}_{\quad \mu}$; it 
can involve completely different tensor fields. General tensor structures as such, however, add no new physical insight in that
their contributions combine with those from $Q^{\beta\nu\alpha}_{\quad \mu}$ in forming the gravitational field equations. 
Thus, one can regard (\ref{soln3-riemann}) as a conservative formulation and proceed to determine if it leads to
a self-consistent solution for  $\tilde{c}_i$, and from (\ref{soln3-riemann}) one gets  
\begin{eqnarray}
\label{soln4-riemann}
{\mathfrak{R}}^{\mu}_{\alpha\nu\beta}\left({}^{g}\Gamma\right) &=& {\mathfrak{W}}^{\mu}_{\alpha\nu\beta} + 
\tilde{\Lambda} \left(\delta^{\mu}_{\nu} g_{\alpha\beta} - g_{\alpha\nu} \delta^{\mu}_{\beta}\right) +
Q^{\mu}_{\alpha\nu\beta}
+ \tilde{c}_1 \big( Q^{\mu}_{\nu} g_{\alpha\beta} - \delta^{\mu}_{\beta} Q_{\alpha\nu}\big)\nonumber\\
&+& \tilde{c}_2 \big( \delta^{\mu}_{\nu} Q_{\alpha\beta} - Q^{\mu}_{\beta} g_{\alpha\nu}\big)
+ \tilde{c}_3  Q \big( \delta^{\mu}_{\nu} g_{\alpha\beta} - \delta^{\mu}_{\beta} g_{\alpha\nu}\big)
\end{eqnarray}
as a relation between the Riemann tensor and the non-geometrical tensor field $Q^{\mu}_{\alpha\nu\beta}$. Is
this an equation of motion for the Riemann tensor? No, there is no such thing. The reason is that, in the presence of 
a metric tensor (see the earlier work \cite{non-lin2} for non-metrical setup) that defines the 
Levi-Civita connection ${}^{g}\Gamma$, the Riemann tensor decomposes into Weyl and Ricci parts 
\begin{eqnarray}
\label{soln4p-riemann}
{\mathfrak{R}}^{\mu}_{\alpha\nu\beta}\left({}^{g}\Gamma\right) &=& {\mathfrak{W}}^{\mu}_{\alpha\nu\beta}\left({}^{g}\Gamma\right) 
+ \frac{1}{2} \big( \delta^{\mu}_{\nu} {\mathcal{R}}_{\alpha\beta}\left({}^{g}\Gamma\right) - {\mathcal{R}}^{\mu}_{\beta}\left({}^{g}\Gamma\right) g_{\alpha\nu}\big)\nonumber\\
&+& \frac{1}{2} \big( {\mathcal{R}}^{\mu}_{\nu}\left({}^{g}\Gamma\right) g_{\alpha\beta} - \delta^{\mu}_{\beta} {\mathcal{R}}_{\alpha\nu}\left({}^{g}\Gamma\right)\big)
- \frac{1}{6} {\mathcal{R}}\left(g, {}^{g}\Gamma\right) \big( \delta^{\mu}_{\nu} g_{\alpha\beta} - \delta^{\mu}_{\beta} g_{\alpha\nu}\big)
\end{eqnarray}
to facilitate a dynamical equation for Ricci tensor. In this sense, the relation (\ref{soln4-riemann}) is 
a rank-4 transcription of the Einstein field equations. It rests wholly on the dynamics of the Ricci 
tensor. Clearly, the Ricci tensor can be determined either by contracting the Riemann tensor in (\ref{soln4-riemann}) 
\begin{eqnarray}
\label{soln4pp-riemann}
{\mathcal{R}}_{\alpha\beta}\left({}^{g}\Gamma\right) &=& {3}\tilde{\Lambda} g_{\alpha\beta} +
(1 - \tilde{c}_1 + 3 \tilde{c}_2) Q_{\alpha\beta} + (\tilde{c}_1 + 3\tilde{c}_3) Q g_{\alpha\beta}
\end{eqnarray}
or by solving for the inverse Ricci piece in (\ref{soln-riemann}) 
\begin{eqnarray}
\label{soln5-riemann}
{\mathcal{R}}_{\alpha\beta}\left({}^{g}\Gamma\right) &=& \frac{1}{a}\left(\Lambda- 2 b \tilde{\Lambda}\right) g_{\alpha\beta}
+ \frac{2 }{3 a} \left[(1 + 3 c_1 - c_2) - b (1 +3 \tilde{c}_1 - \tilde{c}_2)\right] Q_{\alpha\beta} \nonumber\\
&+& \frac{2 }{3 a} \left[(c_2 + 3 c_3) - b (\tilde{c}_2 +3 \tilde{c}_3)\right] Q g_{\alpha\beta} 
\end{eqnarray}
where the two solution must agree with each other. Consequently, imposing also the Bianchi identity, the Ricci tensor is found to 
read as 
\begin{eqnarray}
\label{solution-riemann-einstein}
{\mathcal{R}}_{\alpha\beta}\left({}^{g}\Gamma\right) = \frac{3}{3 a + 2 b} \Lambda g_{\alpha\beta} - \frac{\left(2 + 4 c_1\right)}{(3 a + 2 b)}
\left( K_{\alpha\beta} - \frac{1}{2} K g_{\alpha\beta}\right)
\end{eqnarray}
provided that the unknown couplings in the ansatz (\ref{soln3-riemann}) take the specific values 
\begin{eqnarray}
\label{coeffs-riemann}
\tilde{\Lambda} &=& \frac{\Lambda}{3 a + 2 b}\nonumber\\
\tilde{c}_1 &=& \frac{1}{b (12 a + 8 b)} \left( 3 a (3 - 2 b  + 6 c_1)  + 2 b (1 - 2 b - 2 c_1)\right)\nonumber\\
\tilde{c}_2 &=& \frac{1}{b (12 a + 8 b)} \left(- 3 a ( 1 - 2 c_1) - 2 b (3 + 2 c_1) \right)\nonumber\\
\tilde{c}_3 &=& \frac{1}{b (12 a + 8 b)(c_1-1)} \left( a (1 + 5 c_1 - 6 c_1^2) - 2 b (1  + c_1 + 2 c_1^2)\right)
\end{eqnarray}
where $c_1 \neq 1$ as required by (\ref{diverg}), (\ref{coeffs}) and (\ref{Q-tensor}). Having determined $\tilde{\Lambda}$ and $\tilde{c}_i$, 
the ansatz for Riemann tensor in (\ref{soln3-riemann}) gets fully fixed in terms of $\Lambda$, $c_1$ and $Q^{\beta\nu\alpha}_{\quad \mu}$.

The solution for Ricci tensor in (\ref{solution-riemann-einstein}) is the core dynamical equation governing the dynamics of 
the Ricci tensor. Its exegesis as gravitational field equations is dictated by the model parameters $a$, $b$ 
and the motion integrals $\Lambda$, $Q^{\beta\nu\alpha}_{\quad \mu}$. This exegesis phase or reinterpretation stage is critically 
important for making sense of the Riemanned Eddington theory, which is dissected below systematically for 
clarifying its structure in comparison to the GR.
\begin{enumerate}
\item We first study the model parameters. The equation (\ref{solution-riemann-einstein})
contains three dimensionless parameters: $a$, $b$ and $c_1$. Here, $a$ and $b$ are fundamental
constants appearing in the governing action (\ref{ac-riemann}). They both are 
plain constants (not invariants) because only tensor densities can be defined in affine 
spacetime. The parameter $c_1$ is neither a fundamental constant nor a motion integral; 
it is just a parameter that decorates (\ref{solution-riemann-einstein}) to form a one-parameter 
family of solutions for the Ricci tensor. Excepting $c_1 = 1$ and $c_1 = \infty$, it can take any value, 
including $c_1 = 0$. 

The model parameters cannot take arbitrary values. Their ranges must ensure that 
the Eddington theory is transcended to involve non-geometrical sources like $K_{\alpha\beta}$. 
This is crucial because, physically, the Riemanned Eddington theory should reduce to the original Eddington 
theory not because of some parameter takes a bad value but because non-geometrical sources like $K_{\alpha\beta}$
vanish. Below we determine the admissible ranges of model parameters:  
\begin{enumerate}
\item The $b \rightarrow 0$ limit. In this case, as follows from (\ref{coeffs-riemann}), all three $\tilde{c}_i \rightarrow \infty$  ($i=1,2,3$) 
and it becomes obligatory to set $Q^{\mu}_{\alpha\nu\beta} \rightarrow 0$ to keep the Riemann tensor in (\ref{soln4-riemann}) finite. 
This, however, reduces the whole dynamics to Eddington setup because both (\ref{soln4-riemann}) and (\ref{solution-riemann-einstein}) 
give the Eddington solution (\ref{solution0}). In accordance with the action (\ref{ac-riemann}), $b\rightarrow 0$ limit necessarily 
yields the Eddington result.

\item The $a \rightarrow 0$ limit.  In this limit, the Riemann piece itself is forced to satisfy the equation of motion (\ref{eom-riemann}), and Bianchi
identities require $c_1 = -{1}/{2}$. However, this particular $c_1$ value kills the $K_{\alpha\beta}$ contribution in 
(\ref{solution-riemann-einstein}), leaving behind precisely the Eddington solution (\ref{solution0}). 

\item The $c_1 \rightarrow -{1}/{2}$ limit. This limit is reached through also $a\rightarrow 0$ limit, as described just above. It
completely erases the matter part of (\ref{solution-riemann-einstein}), leaving behind again the Eddington solution.
\end{enumerate}
In consequence, Eddington theory is transcended to incorporate non-geometrical sources like $K_{\alpha\beta}$ if 
$a\neq 0$ and $b\neq 0$ and $c_1 \neq -1/2$ simultaneously, and it is attained back only when $K_{\alpha\beta}$ itself vanishes. 

\item Having dealt with the model parameters, we now turn to a detailed discussion of the motion integrals $\Lambda$ and $K_{\alpha\beta}$. 
In general, they fall in two physically distinct categories depending on whether they are independent or not, and the gravitational field
equations they lead to read as follows.   
\begin{enumerate}
\item {\it $\Lambda$ and $K_{\alpha\beta}$ are independent quantities.} By this we mean that, they are related neither parametrically nor dynamically. 
In this case, in (\ref{solution-riemann-einstein}), one is free to perform the rescalings $\displaystyle{\frac{3\Lambda}{(3 a + 2 b)} \rightarrow \Lambda}$ 
and $\displaystyle{-\frac{(2 + 4 c_1)}{(3 a + 2 b)} K_{\alpha\beta} \rightarrow K_{\alpha\beta}}$ to eliminate $a$, $b$, $c_1$ from equations. This simplifies 
the right-hand side of (\ref{solution-riemann-einstein}) to  $\Lambda g_{\alpha\beta} + K_{\alpha\beta}$, in which while the homogeneous part $\Lambda$ sets 
the energy-momentum tensor for empty space
\begin{eqnarray}
\label{en-mom-vac}
T^{(vac)}_{\alpha\beta} = - M_{Pl}^2 \Lambda g_{\alpha\beta}
\end{eqnarray}
the non-geometrical part $K_{\alpha\beta}$ sets the energy-momentum tensor of matter
\begin{eqnarray}
\label{en-mom-mat}
T^{(matt)}_{\alpha\beta} = M_{Pl}^2 {K_{\alpha\beta}}
\end{eqnarray}
with $M_{Pl}^2 = (8\pi G_N)^{-1/2}$ being the fundamental scale of gravity. These associations between curvature sources 
and energy-momentum tensors make sense as long as the two energy-momentum tensors do not convert into each other. 
In other words, material system must be genuinely classical and must undergo no phase transition. This is because phase 
transitions induce constant curvatures that add to $\Lambda$ -- an effect that makes $\Lambda$ to depend explicitly on the 
matter sector parameters whereby disrupting independent rescalings of $\Lambda$ and $K_{\alpha\beta}$. Consequently,
the model parameters $a$, $b$ and $c_1$ can be embedded into motion integrals to obtain exactly the gravitational 
field equations in GR provided that the energy-momentum tensors of matter and empty space are independent each other.
And this can happen particularly in classical systems undergoing no phase transitions.  

\item {\it $\Lambda$ and $K_{\alpha\beta}$ are not independent quantities.} By this we mean that, they can have common parameters. 
In fact, all realistic systems fall in this category. Now, because of their mutual dependence, $\Lambda$ and $K_{\alpha\beta}$
cannot be rescaled freely simply because the rescaled $\Lambda$ cannot be guaranteed to correspond to the $\Lambda$ which would 
be induced by the rescaled $K_{\alpha\beta}$. In general, $\Lambda$ and $K_{\alpha\beta}$ develop nontrivial correlations 
through phase transitions and quantum corrections because these effects give contributions to $\Lambda$ in a way 
explicitly involving the matter sector parameters along with the momentum cutoffs applied \cite{quantum-corr}. Consequently, the vacuum energy-momentum 
tensor in (\ref{en-mom-vac}) and the matter energy-momentum tensor in (\ref{en-mom-mat}) cannot be rescaled independently 
to eliminate the parameters $a$, $b$ and $c_1$ in (\ref{solution-riemann-einstein}). This means that the Ricci tensor there  wholly sets the sought gravitational field equations 
\begin{eqnarray}
\label{eq-einstein}
{\mathcal{R}}_{\alpha\beta}\left({}^{g}\Gamma\right) =  \frac{1}{M_{Pl}^2}  \left( T_{\alpha\beta} - \frac{1}{2} T g_{\alpha\beta}\right)
\end{eqnarray}
wherein the energy-momentum tensor
\begin{eqnarray}
\label{en-mom}
T_{\alpha\beta} = \frac{3}{(3 a + 2 b)} T^{(vac)}_{\alpha\beta}  - \frac{(2 + 4 c_1)}{(3 a + 2 b)} T^{(matt)}_{\alpha\beta}
\end{eqnarray}
combines the vacuum and matter energy-momentum tensors with $a$, $b$ and $c_1$ staying as additional parameters not found in GR. 
In the language of effective field theory, one first constructs quantum effective action by integrating out all quantum 
fluctuations beyond the matching scale and then extracts from it $T^{(vac)}_{\alpha\beta}$ and $T^{(mat)}_{\alpha\beta}$ \cite{quantum-corr}.
They are thus intimately correlated. The way they contribute to (\ref{en-mom}) involves not just $1/M_{Pl}^2$ as in GR 
but also $a$, $b$ and $c_1$ as extra degrees of freedom. In this sense, vacuum and matter can be made to gravitate differently 
-- a novel feature not found in GR. Nevertheless, if matter is to gravitate as experimented and observed it is necessary to set
\begin{eqnarray}
\label{empirical}
3 a + 2 b = -(2 + 4 c_1)
\end{eqnarray}
so that the material part of (\ref{en-mom}) becomes $+T^{(matt)}_{\alpha\beta}/M_{Pl}^2$. This is precisely the 
way matter gravitates in the GR.

The empirical condition in (\ref{empirical}) calibrates weight of matter. It, however, leaves weight of vacuum undetermined. In exact GR limit, 
one sets $3 a + 2 b = 3$ so that cosmological constant $\Lambda$ gravitates with the same strength as matter. Then, the gravitational field equation 
(\ref{eq-einstein}) gives a maximally symmetric background geometry with ${\mathcal{R}}\left(g, {}^{g}\Gamma\right) = 4 \Lambda$. The cosmological 
observations \cite{obs} require this curvature to acquire a tiny value ${\mathcal{R}}^{(obs)} \sim m_{\nu}^4/M_{Pl}^2$, $m_{\nu}$ being the neutrino mass scale.
This vanishingly small curvature,  given the diversity and enormity of contributions to vacuum energy \cite{ccp}, is perplexingly difficult to realize. 
This is the well-known cosmological constant problem \cite{ccp} -- the severest naturalness problem plaguing both particle physics and cosmology. 

It is clear that, cosmological constant problem is special to GR because it is in GR that a bound on curvature directly translates 
into a bound on $\Lambda$. Fortunately, in the Riemann-improved Eddington theory being developed, this impasse metamorphoses into a 
tractable problem simply because there is no physical reason as in (\ref{empirical}) that enforces $3 a + 2 b = 3$ (or $c_1=-5/4$). As a matter of fact, 
these fundamental constants must be fixed by using observational data on the cosmological constant itself \cite{obs}. Then, extreme smallness of ${\mathcal{R}}^{(obs)}$ 
necessitates at least one of $a$, $b$ to be a large number. Pertaining to a physical theory, in the philosophy of Dirac's large number hypothesis, 
$a$ and $b$ can be interpreted as the ratio of two hierarchically different mass scales. By force of the gravitational nature of the setup, 
one of the mass scales is $M_{Pl}$. The other one, denoted hereon by $M_{Co}$, is plausibly a cosmological scale. Letting $a\propto (M_{Co}/M_{Pl})^{n_{a}}$, 
$b\propto (M_{Co}/M_{Pl})^{n_{b}}$, and fixing their proportionality constants appropriately one can write
\begin{eqnarray}
\label{a ve b}
a + \frac{2}{3} b =  \left(\frac{M_{Co}}{M_{Pl}}\right)^{n}
\end{eqnarray}
where $n = {\mbox{max}}\left\{ n_a, n_b\right\}$. Replacement of this parametrization in  (\ref{eq-einstein}) gives ${\mathcal{R}}\left(g, {}^{g}\Gamma\right)  = 4 \Lambda \left(M_{Pl}/M_{Co}\right)^2$ 
for $n = 2$. This prediction complies with existing bounds \cite{obs} without constraining $\Lambda$ itself provided that it is possible to realize
\begin{eqnarray}
\label{MCo}
\left(M^{(obs)}_{Co}\right)^2 \simeq \Lambda \left(\frac{M_{Pl}}{m_{\nu}}\right)^{4}
\end{eqnarray}
for $\Lambda > 0$ (signs of $a$ and/or $b$ are reversed for $\Lambda < 0$). For instance,  for $\Lambda \sim M_{Pl}^2$ -- the worst case \cite{ccp} in GR --, 
one gets $M^{(obs)}_{Co} \sim M_{Pl} (M_{Pl}/m_{\nu})^2 \sim 10^{80}\ {\rm GeV}$. This is the mass of the Universe (which consists of some $10^{80}$ hydrogen atoms). 
In general, smaller the $\Lambda$ closer the $M_{Co}$ to $M_{Pl}$. It is astonishing that this whole procedure involves no fine-tuning. 
Indeed, the observed curvature ${\mathcal{R}}^{(obs)} \sim m_{\nu}^4/M_{Pl}^2$ is reproduced not by finely tuning parameters but 
by hierarchically splitting $M_{Co}^2$ and $\Lambda$ via (\ref{MCo}). Consequently, the Riemann-improved Eddington theory , or simply the Riemanned Eddington theory, 
proves firmly natural. It is able to cover the 120 orders of magnitude discrepancy between the observed curvature ${\mathcal{R}}^{(obs)} \sim m_{\nu}^4/M_{Pl}^2$ \cite{obs} 
and the theoretical expectation ${\mathcal{R}}^{(thr)} \sim M_{Pl}^2$ \cite{ccp} without enforcing any fine adjustments of model parameters.  
This crucial feature is manifestly reflected in the refined gravitational field equations
\begin{eqnarray}
\label{eq-einstein-2}
{\mathcal{R}}_{\alpha\beta}\left({}^{g}\Gamma\right) =  \frac{1}{M_{Co}^2} \Big(V_0 g_{\alpha\beta}\Big)
+ \frac{1}{M_{Pl}^2} \left( T^{(matt)}_{\alpha\beta} - \frac{1}{2} T^{(matt)} g_{\alpha\beta}\right)
\end{eqnarray}
following from (\ref{eq-einstein}) after imposing (\ref{en-mom-mat}), (\ref{empirical}), (\ref{a ve b}), and
defining the vacuum energy $V_0 = M_{Pl}^2 \Lambda$. This dynamics ensures that, in the Riemanned 
Eddington theory matter and vacuum gravitate not universally with Newton's constant but individually 
with their own gravitational constants. In particular, the vacuum energy $V_0$ gravitates not with Newtonian 
strength $M_{Pl}^{-2}$ but with cosmological strength $M_{Co}^{-2}$. It is this difference between
gravitational constants that enables a local observer to distinguish between vacuum energy and temporary 
blimps in matter energy-momentum tensor. Consequently, in contrast to non-local, acausal modifications
of Einstein gravity which degravitate cosmological constant like (\ref{eq-einstein-2}) without 
fine-tuning \cite{cosmo1, cosmo2}, the Riemann-improved Eddington theory facilitates a local, 
causal and natural setup that degravitates the cosmological constant.

It is timely to ask: Does the Rieamann-extended Eddington theory solve the cosmological constant problem? 
No! There is no such thing. The reason is that the cosmological scale $M_{Co}$ (equivalently, the fundamental constants $a$ 
and $b$) must be a rather specific function of $\Lambda$, as given in (\ref{MCo}). The model, however, offers no dynamics 
for such a functional relation. The relation (\ref{MCo}) is an empirical relation not a fundamental one derived from dynamics.
In fact, $M_{Co}$ is a fundamental constant yet $\Lambda$ is a motion integral, and the two
do not have to have a predefined relation like (\ref{MCo}). Thus, degravitation of the cosmological constant in Riemanned
Eddington theory is local, causal and not fine-tuned; however, the gravitational constant of the vacuum, $M_{Co}$, cannot be fixed as 
a fundamental constant unless $\Lambda$ is known to take a fixed, predefined value. In other words, $\Lambda$ must be known
in terms of the model parameters (gravitational constant, particle masses and the like) as a universal quantity. More specifically, given matter
spectrum and underlying dynamics, one must be able to compute $\Lambda$ to fix $M_{Co}$ in the fundamental action (\ref{ac-riemann}). 
Could this be done? Yes and no. The answer is  yes because the zero-point energies of quantum fields (see \cite{ccp} 
as well as explicit predictions in \cite{quantum-corr}) give a vacuum energy density $V_0 \simeq k_{UV}^4$, $k_{UV}$ being 
the ultraviolet momentum cutoff. The answer is no because it is possible to know neither $k_{UV}$ nor particle spectrum 
beyond the Fermi scale. It is with certain plausible assumptions that one can have an idea about $\Lambda$. For instance,
if supersymmetry is not a symmetry of Nature at short distances, one necessarily takes  $k_{UV}\simeq M_{Pl}$ 
as the ultimate ultraviolet scale for any quantum field theory, and the resultant vacuum energy does not curve 
spacetime overly if $M_{Co}$ is set in (\ref{ac-riemann}) as the mass of the Universe. Summarizing,
having no information about the physics beyond the Fermi scale, it is not possible to determine $\Lambda$ from
the outset, and hence, the relation (\ref{MCo}) cannot facilitate a solution for the cosmological constant problem.
In view of (\ref{eq-einstein-2}), one immediately notes that all that $M_{Co}$ can do is to provide a 
rationale, not a mechanism, for degravitating the cosmological constant.
\end{enumerate}
\end{enumerate}

We close this section by configuring the matter action. Though its energy-momentum 
tensor $T^{(matt)}_{\alpha\beta}$  is the only attribute needed for gravitational field equations, 
the matter action itself is needed for quantization of matter and determination of conserved charges. 
In affine spacetime, however, one cannot write a matter action like (\ref{ac-riemann}) because there is no such determinant
in matter sector \cite{wilczek} (see also \cite{non-lin} in this respect). It is actually not needed at all.
The reason is that the matter action emerges dynamically along with the metric tensor, the Levi-Civita
connection and the energy-momentum tensor. In other  words, once metric and energy-momentum tensor
are induced, the matter action is constructed as that functional of matter fields whose variation
with respect to the metric tensor gives the energy-momentum tensor in (\ref{en-mom}) 
\cite{kirsch,review2,non-lin,non-lin2}. This association of energy-momentum tensor (see \cite{en-mom} 
for a detailed discussion of energy-momentum current) leads to a dynamical equivalence 
$I_{R}\left[\Gamma\right] \leadsto I\left[g,\psi\right]$, where
\begin{eqnarray}
\label{ac-equiv}
I\left[g,\psi\right] = \int d^{4}x \sqrt{-g} \left\{ \frac{1}{2} M_{Pl}^2  g^{\alpha \beta} {\mathcal{R}}_{\alpha\beta}\left({}^{g}\Gamma\right)
- \frac{M_{Pl}^2}{M_{Co}^2} V_0 + {\mathcal{L}}_{(matt)}\left(g,\psi\right) \right\}
\end{eqnarray}
is the matter and gravity action in spacetimes endowed with metric tensor $g_{\alpha\beta}$. It gives
the Einstein field equations (\ref{eq-einstein-2}) through $\displaystyle{\frac{\delta I} {\delta g^{\alpha\beta}} = 0}$
where $\displaystyle{T^{(matt)}_{\alpha\beta} = - 2 \frac{\delta {\mathcal{L}}_{(matt)}}{\delta g^{\alpha\beta}} + g_{\alpha\beta} {\mathcal{L}}_{(matt)}}$
is the matter energy-momentum tensor. The equations of motion of the matter fields $\psi$ follow from $\displaystyle{\frac{\delta I} {\delta \psi} = 0}$,
with ${\mathcal{L}}_{(matt)}\left(g,\psi\right)$ being the matter Lagrangian. The metrical framework of (\ref{ac-equiv}), 
dynamically equivalent to (\ref{ac-riemann}), precisely gives the GR plus matter except for one single term, which is the cosmological constant.
The $\Lambda$ term in GR is replaced by $(\Lambda M_{Pl}^2)/M_{Co}^2$, which can degravitate $\Lambda$ for sufficiently large $M_{Co}$ (or 
$a$ and/or $b$). Moreover, dynamical equivalence on material dynamics ensures that there exists actually no reason to 
expect instabilities in tensor or other modes (They are known to occur when Eddington action (\ref{ac-eddington}) is directly 
added the matter action \cite{banados1,banados2}).

\section{Discussions and Conclusion}
In summary, the findings and implications of the present work are as follows:
\begin{itemize}
\item The Riemann-improved Eddington theory, under reasonable ansatze, gives a dynamical equation for Ricci tensor. In view of its decomposition into Weyl and Schouten tensors, however, the Riemann tensor cannot possess an independent dynamical equation by itself. Gravitational waves are accommodated by the presence of the Weyl curvature in the ansatz for Riemann tensor. 

\item The resulting dynamical equation for Ricci tensor can be put in exact GR form. However, parameter space is wide enough to allow for novel solutions where Ricci curvature is sourced by cosmological constant and matter not directly as in GR but through nontrivial dressings by fundamental constants. Nevertheless, its coupling to the energy-momentum 
tensor of matter can be brought to observed strength by correlating model parameters, and this ensures that matter gravitates just as in the GR.

\item The dressing factor for cosmological constant is set also by observations. 
Interestingly, this determination is accomplished not by finely tuning model parameters 
as in GR but by hierarchically splitting the cosmological constant and the 
cosmological scale introduced by the model. The cosmological scale evaluates 
around the mass of the Universe when the cosmological constant lies at the Planck scale.

\item The cosmological scale must be born in the theory with right value to suppress the
cosmological constant. The model, however, does not offer a dynamical basis for such a relation
so that the said relation is empirical rather than fundamental.

\item The metric and energy-momentum tensors facilitate introduction of an effective
matter action, and this establishes a dynamical equivalence between the Riemanned Eddington 
theory and a GR-like metrical theory of gravity. The latter differs from GR plus matter setup only by the 
degravitation factor in front of the cosmological constant. The vacuum energy receives contributions
from all zero-point energies and phase transitions yet it weighs too small to cause cosmological
constant problem, at least at the empirical level. 

\item It must be kept in mind that the GR-like gravitational dynamics found in the present work is based on 
a specific ansatz which solves the affine equations of motion. There can exist other solutions as well. In
this sense, the Riemanned Eddington theory can give rise to alternative gravitational theories which
may or may not reduce to GR. 
\end{itemize}
The Riemann-improved Eddington theory can be further generalized, for instance, by considering determinants of higher-rank 
curvature tensors, by switching on torsion, or by going to higher-dimensions. In addition, one can 
search for other possible solutions of the affine equations of motion. If any, it would be 
interesting to see if there are specific solutions that can provide a fundamental resolution for the 
cosmological constant problem. Also interesting would be a proper understanding of the 
flat rotation curves of galaxies {\it a la} Milgrom's modified gravitational dynamics 
(cold dark matter being a competing alternative). These attempts, if prove positive, would 
further strengthen Eddington's affine gravity program.

In summary, the Riemann-improved Eddington theory, inherently formulated in affine space, 
leads to a novel framework in which one preserves all the successes of the GR while fathoming
the observed size of the Universe in a local, causal, natural but, alas, empirical way.

\vspace{0.4cm}
\noindent{\bf Acknowledgements:} The work at the ICTP High Energy Section is carried out through the ICTP Associate Programme (August 2013).
I am indebted to conscientious referees for their fruitful comments, suggestions and criticisms.

\end{document}